\documentclass[
preprint,showpacs,preprintnumbers,amsmath,amssymb]{revtex4}
\usepackage{mathrsfs}
\usepackage{graphicx}
\usepackage{epsfig}
\usepackage{dcolumn}
\usepackage{bm}

\let\jnfont=\rm
\def\NPB#1,{{\jnfont Nucl.\ Phys.\ B }{\bf #1},}
\def\PLB#1,{{\jnfont Phys.\ Lett.\ B }{\bf #1},}
\def\EPJC#1,{{\jnfont Euro.\ Phys.\ J.\ C }{\bf #1},}
\def\PRD#1,{{\jnfont Phys.\ Rev.\ D }{\bf #1},}
\def\PRL#1,{{\jnfont Phys.\ Rev.\ Lett.\ }{\bf #1},}
\def\MPLA#1,{{\jnfont Mod.\ Phys.\ Lett.\ A }{\bf #1},}
\def\JPG#1,{{\jnfont J.\ Phys.\ G}{\bf #1},}
\def\CTP#1,{{\jnfont Commun.\ Theor.\ Phys.\ }{\bf #1},}
\def\JHEP#1,{{\jnfont J. High \ Ener.\  Phys.}{bf #1},}
\def\RMP#1,{{\jnfont  Rev. Mod. Phys.}{bf #1},}

\def\E_slash{\not{\hbox{\kern-2.1pt $E$}}}

\begin{document}
\preprint{}

\title{Probing topcolor-assisted technicolor models from like-sign $\tau $ pair
 production in $e\gamma$ collisions }

\author{Guo-Li Liu}
\affiliation{ Physics Department, Zhengzhou University, Henan,
Zhengzhou 450001, China \\
 Kavli Institute for Theoretical Physics China, Academia Sinica,
Beijing 100190, China}
\thanks{E-mail:guoliliu@zzu.edu.cn}

\date{\today}

\begin{abstract}
We consider the contributions of the extra gauge boson $Z'$
to the like sign $\tau$ production process $e^-\gamma\to e^+(\mu^+)\tau^-\tau^-$,
induced by the tree-level flavor changing interactions.
Since these rare production are
far below the observable level in the Standard Model and other popular new physics
 models such as the minimal supersymmetric model, we find that $Z'$ can give significant
 contributions to this process,  and with reasonable values of the parameters in
 TC2 models, the cross section $\sigma$ can reach several tens of fb and may be
detected at the $e\gamma$ collisions.
\end {abstract}

 \maketitle

 The mechanism of electroweak symmetry breaking (EWSB) remains
the most prominent mystery in elementary particle physics. Probing
EWSB will be one of the most important tasks in the future high
energy colliders. Dynamical electroweak symmetry breaking (EWSB),
such as technicolor (TC) theory\cite{tc-theory}, is an attractive
idea that it avoids the shortcomings of triviality and unnaturalness
arising from the elementary Higgs field. TC2 theory is an attractive
scheme in which there is an explicit dynamical mechanism for
breaking electroweak symmetry and generating the fermion masses
including the heavy top quark mass. It is one of the important
promising candidates for the mechanism of EWSB.

In TC2 theory \cite{tc2-theory},  EWSB is driven mainly by TC
interactions, the extended technicolor (ETC) interactions give
contributions to all ordinary quark and lepton masses including a
very small portion of the top quark mass, namely
$m_{t}^{\prime}=\epsilon m_t$ with a model-dependent parameter
$\epsilon(\epsilon\ll 1)$. The topcolor interactions also make small
contributions to EWSB and give rise to the main part of the top
quark mass $m_t-m_{t}^{\prime}=(1-\epsilon)m_t$ similar to the
constituent masses of the light quarks in QCD. This means that the
associated $Z'$ is the physically observable objects. Thus $Z'$ can
be seen as the characteristic feature of TC2 theory. Studying the
possible signatures of $Z'$ at future high energy colliders can be
used to test TC2 theory and further probe the EWSB mechanism.

The third generation is treated differently, which is the
characteristic feature of the TC2 theory. As the heavies lepton, the
properties of the $\tau$ lepton are distinctive --it may have larger
tree-level flavor changing couplings, such as $\tau-\mu$ and
$\tau-e$ transformation. At the same time, there are many kinds of
new physics scenarios predicting new particles, which can lead to
significant LFC signals. For example, in the minimal supersymmetric
standard model(SM), a large $\nu_\mu-\nu_\tau$ mixing leads to clear
LFC signals in slepton and lepton collider\cite{vv-mixing}. The
non-universal U(1) gauge bosons $Z'$, which are predicted by various
specific models beyond the SM, can lead to the large tree-level
flavor changing(FC) couplings. Thus, these new particles may have
significant contributions to some LFC processes\cite{z'couple}.

The international linear collider(ILC) offers excellent new
opportunities for the study of high energy particle collisions. The
idea to convert the electron
 beams of a ILC into photon beams, by laser backscattering, and thus create a
 photon collider, was first discussed almost 30 years ago in\cite{photon collider},
  and then studied sufficiently in the coming years\cite{er-reviews}. With the
  luminosity and energy of such colliders being comparable to those of the basic
   $e^+e^-$ collider, one may now consider the process such as $e^- \gamma \to
   e^+(\mu^+)\tau^-\tau^-$.

Some leptonic flavor violations in the presence of an extra Z' has
been studied in the literature\cite{z'-process}. In this note, we
calculate the contributions of the extra $U(1)$ gauge boson $Z'$
 to the flavor violating process $e^-\gamma\to e^+(\mu^+)\tau^-\tau^-$
and see whether $Z'$ can be detected via this process at high-energy
linear $e\gamma$ collision experiments. We find that this process is
important in probing the gauge boson $Z'$. With reasonable values of
the parameters in TC2 models, the signal rates can be fairly large,
which may be detected at the $e\gamma$ colliders based on the ILC
experiments.

  For TC2 models \cite{tc2-theory},
the underlying interactions, topcolor interactions, are
non-universal and therefore do not posses a GIM mechanism. This is
an essential feature of this kind of models due to the need to
single out the top quark for condensation. This non-universal gauge
interactions result in the FC coupling vertices when one writes the
interactions in the quark mass eigenbasis. Thus the extra gauge
boson predicted by this kind of models have large couplings to the
third generation and can induce the FC couplings.

The couplings of the extra $U(1)$ gauge bosons $Z'$ to the ordinary
fermions can be written as \cite{exp-tc2}:
\begin{equation}
{\cal L}=-\frac{1}{2}g_{1}\{K_{\mu e}(\bar{e}_{L}\gamma^{\mu}
         \mu_{L}+2\bar{e}_{R}\gamma^{\mu}\mu_{R})+k_{\tau\mu}(\bar{\tau}_{L}
         \gamma^{\mu}\mu_{L}+2\bar{\tau}_{R}\gamma^{\mu}\mu_{R})
+k_{\tau e}(\bar{\tau}_{L}
         \gamma^{\mu}e_{L}+2\bar{\tau}_{R}\gamma^{\mu}e_{R})  \}\cdot Z'_{\mu},
         \label{coup-1}
\end{equation}

where $g_{1}$ is the ordinary hypercharge gauge coupling constant
and $k_{\mu e}$, $k_{\tau e}$ and $k_{\tau\mu}$ are the flavor
mixing
 factors. Since the new gauge boson $Z'$ couples preferentially to the third
 generation, the factor $K_{\mu e}$ are negligibly small, so in the
 following estimation, we will neglect the $\mu- e$ mixing, and consider only
 the flavor changing coupling processes $e\gamma \to \bar e(\bar\mu)\tau\tau$.

Note that the difference between the $Z'\tau\bar\mu$ and
$Z'\tau\bar
 e$ couplings lies only in the flavor mixing factor $K_{\tau\mu}$
and $K_{\tau e}$ and the masses of the final state $\mu$ and $e$ leptons. Since the
non-universal gauge boson $Z'$ treats the fermions in the third generation
differently from those in the first and second generations and treats the
 fermions in the first same as those in the second generation, so in the
following calculation, we will assume $K_{\tau\mu} = K_{\tau e}$.
Then what makes the discrepancy of the cross
sections of the two channels $e\gamma \to \bar e\tau\tau$ and
$e\gamma \to \bar\mu\tau\tau$ is only the masses the final state
particles. Considering the large mass of the $Z'$,  $M_{Z'}> 1 TeV$,
for simplicity, We will take $m_\mu=m_e=0$ in the
following discussion, i.e., assuming the cross sections of the two channels
$e\gamma \to \bar e\tau\tau$  and $e\gamma \to \bar\mu\tau\tau$  are equal to
 each other and take the former as an example in the following discussion.

The TC2 parameters concerned in this process are $K_{\tau e}$,
$K_{\tau\mu}$, $K_{e\mu}$, $K_1$ and the mass of the extra gauge
boson $M_Z'$. $K_{e\mu}$ is very small, about $10^{-3}$, we will not
consider the process induced by the coupling with it.  In our
calculation, we have assumed $K_{\tau\mu}=K_{\tau e}  = K_{\tau l}$
($l=e,\mu$)\cite{exp-tc2,tc2-cla}. In fact, for the $TC2$ models,
the extended gauge groups are broken at the $TeV$ scale, which
proposes that $K_{\tau l}$ is an $\cal{O}$$(1)$ free parameter. Its
value can be generally constrained by the current experimental upper
limits on the $LFV$ processes $l_{i}\rightarrow l_{j}\gamma$ and
$l_{i}\rightarrow l_{j}l_{k}l_{l}$. However, from the numerical
results of Ref.\cite{t-eu}, we can see that the $LFV$ processes
$l_{i}\rightarrow l_{j}\gamma$ and $l_{i}\rightarrow
l_{j}l_{k}l_{l}$ can not give severe constraints on the mixing
factor $K_{\tau l}$. Thus, in our calculation, we choose $K_{\tau
l}$ in the range of $0-1$, which is expected consistent with
theoretically-allowed parameter regions and also with current
experimental data.

 It has been shown
that the vacuum tilting (the topcolor interactions only condense the
top quark but not the bottom quark), the coupling constant $K_{1}$
should satisfy certain constraint, i.e. $K_{1}\leq 1$ \cite{K_1}. We
choose $K_1 = 0.2$ since the $K_1$ occures only in the decay width
of $Z'$ and affects the cross section slightly.

  The lower limits on the mass $M_Z'$ of the new gauge boson $Z'$
predicted by topcolor $Z'$ scenario can be obtained via studying its
effects on various observable, which has been extensively
studied\cite{exp-tc2}. For example, Ref.\cite{EW-data} has shown
that, to fit the electroweak precision measurement data, the $Z'$
mass $M_Z'$ must be larger than $1$ TeV. The lower Z' bounds on
$M_Z'$ can also be obtained from dijet and dilepton production at
the Tevatron $Z'$ experiments\cite{z-mass-dijet-dilepton}, or from
$B\bar B$ mixing\cite{bb-mixing}. However, these bounds are
significantly weaker than those from precisely electroweak data.
Furthermore, Refs.\cite{k1-z-mass} have shown that, for the coupling
parameter $K_1 < 1$, the $Z'$ mass $M_Z'$ can be explored up to
several TeV at the ILC experiment with $\sqrt{S} =500$ GeV and the
integrated luminosity $L_{int} =100$ $fb^{-1}$ . As numerical
estimation, we will take $M_Z'$ as a free parameter and assume that
$M_Z'$ is in the range of $1$TeV - $2.5$TeV throughout this paper.
Finally, Note that the charge conjugate $\bar\tau\bar \tau \mu(e)$
production channel are also included in our numerical study.

 The total decay width of the extra gauge boson $Z'$ is dominated,
 since the topcolor scenarios treat the third generation differently,
by the third generation, i.e., the $t\bar t$, $b\bar b$, $\tau\bar \tau$
 and the $\nu_\tau\bar \nu_\tau$ channels, which can be approximately
calculated as:
\begin{eqnarray}
\Gamma_Z' \sim \frac{g_1^2 cot^\theta}{4
\pi}M_Z'(\frac{5}{4}+\frac{1}{3}) \sim K_1M_Z'
\end{eqnarray}
Where the former factor in the bracket is from the lepton
contribution, while the quarks give the latter result.

 \begin{figure}[tbh]
\begin{center}
\epsfig{file=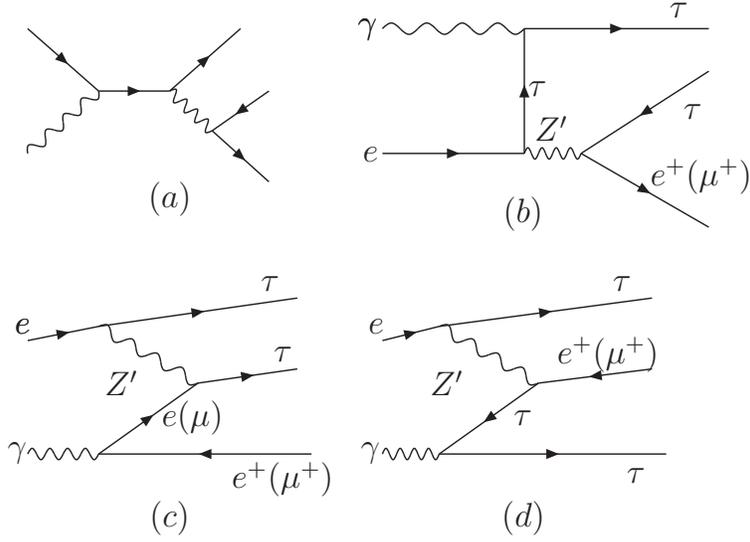,width=10cm} \caption{ Feynman diagrams
contributing to the process $e\gamma \to e^+(\mu^+)\tau\tau$ in TC2
models.} \label{fig1}
\end{center}
\end{figure}

The Feynman diagram for the $Z'$ gauge boson contributions to the
process $e^-\gamma\to e^+\tau^-\tau^-$ is shown in Fig.1 at the tree
level, from which, we can see there involve, in each diagram, two
LFV vertexes which are strongly depressed in the SM. With Eqn.
\ref{coup-1},
 we can write directly the contribution of $Z'$ to the amplitude
of the process $e^-\gamma\to e^+\tau^-\tau^-$:
\begin{eqnarray}
{\cal M}&=&\frac{1}{4}ieg^2K^2_{\tau\mu}[a_1\bar u_{\tau,2} \gamma_\nu(P_L+2P_R)v_{e2} \cdot \bar u_{\tau,1} \gamma_\nu (P_L+2P_R)(p_{e1}+p_\gamma)\gamma_\mu u_e \epsilon_{\mu} \nonumber \\
&+& a_2\bar u_{\tau,2} \gamma_\nu(P_L+2P_R)v_{e2} \cdot \bar u_{\tau,1} \gamma_\mu (p_{\tau,2}-p_\gamma)\gamma_\nu(P_L+2P_R) u_e \epsilon_{\mu}   \nonumber \\
&+& a_3 \bar u_{\tau,2} \gamma_\nu(P_L+2P_R)p_\gamma\gamma_\mu v_{e2} \cdot \bar u_{\tau,1} \gamma_\nu (P_L+2P_R) u_e \epsilon_{\mu}  \nonumber \\
&+& a_4\bar u_{\tau,2} \gamma_\mu(p_{\tau,2}-p_\gamma)\gamma_\mu
v_{e2} \cdot \bar u_{\tau,1} \gamma_\nu (P_L+2P_R)
u_e\epsilon_{\mu}] \label{amp}
\end{eqnarray}
The expressions $a_1$, $a_2$, $a_3$ and $a_4$ in equation.\ref{amp} are
  given as,
 \begin{eqnarray}
 a_1&=& \frac{1}{(p_e+p_\gamma)^2}\frac{1}{(p_{\tau 2}-p_{\bar e})^2-M_{Z'}^2 },  \nonumber \\
 a_2&=&\frac{1}{ (p_{\tau 1}-p_\gamma)^2}\frac{1}{ (p_{\tau 2}+p_{\bar e})^2-M_{Z'}^2   }, \nonumber \\
 a_3&=&  - \frac{1}{(p_{\bar e}- p_\gamma)^2}\frac{1}{(p_{\tau 1}- p_e)^2 -M_{Z'}^2 },  \nonumber \\
 a_4&=& \frac{1}{(p_{\tau 2} - p_\gamma)^2}\frac{1}{(p_{\tau 1}- p_e)^2 -M_{Z'}^2 }.
\end{eqnarray}
 Where $p_{e}$($p_{\bar{e}})$ denotes the momentum of the initial $e$
(the final state $\bar e$), $p_{\tau 1}$ and $p_{\tau 2}$ denotes
the momenta of the final two like-sign $\tau$ particles  and
$p_\gamma$, the initial photon momentum; $P_{R,L}=(1\pm\gamma^5)/2$
are the chiral operator.

The hard photon beam of the $e^-\gamma$ colliders can be obtained
from laser backscattering at the ILC \cite{er-collision}. We define
that $\sqrt{\hat{s}}$ and $\sqrt{s}$ are the center-of-mass energies
of the $e^-\gamma$ and $e^+e^-$ colliders, respectively. After
calculating the cross section $\sigma(\hat{s})$ for the subprocess
$e^-\gamma\to e^+(\mu^+)\tau\tau$, the total cross section
$\sqrt{s}$ at the ILC experiments can be obtained by folding
$\sigma(\hat{s})$ with the backscattered laser photon spectrum
$f_{\gamma}(x) (\hat{s}=x^2s)$
 \begin{equation}
 \sigma = \int_{2m_t/\sqrt{s}}^{x_{max}}dx\hat{\sigma}
 (\hat{s})f_{\gamma}(x).
\end{equation}
The backscattered laser photon spectrum $f_{\gamma}(x)$ is given
in Ref.\cite{er-collision}. Beyond a certain laser energy $e^+e^-$ pairs are
produced, which significantly degrades the photon beam. This leads
to a maximum $e\gamma$ centre of mass energy of $\sim 0.91\times
\sqrt{s}$.

In our calculation, we restrict the angles of the observed particles
relative to the beam, $\theta_{e^-}$ and $\theta_{e^+}$ to the range
$10^\circ\leq\theta_{e^-}$, $\theta_{e^+}\leq170^\circ$. We further
restrict the particle energy $E_e\geq 10$ GeV. For simplicity, we
have ignored the possible polarization for the electron and photon
beams. To obtain numerical results, we take $m_\tau=1.777$ GeV,
$m_\mu=0.12$ GeV and $\alpha_e=1/128$ \cite{alpha}. For estimating
the number of the $e^+\tau\tau$
 event, we consider the $e^+e^-$
centre-of-mass energy $\sqrt{s}$ in the range of $300$GeV-$1500$GeV
 appropriate to the TESLA/NLC/JLC high energy colliders and assume an
 integrated luminosity of $L=500 fb^{-1}$.

In Fig.2, we show the cross section $\sigma$ of the process
$e^-\gamma\to e^+(\mu^+)\tau^-\tau^-$ as a function of the mass of
the Z' for three values of the center-of-mass energy $\sqrt{s}$. One
can see that Z' can give significant contributions to the process
$e^-\gamma\to e^+\tau\tau$, and the cross section $\sigma$ is
sensitive to the parameter space. The $Z'$ contribution increases
with the increasing $\sqrt{s}$.

The signature of $ e\gamma \to \bar e (\bar\mu)\tau \tau $ can be
chosen as two like-sign leptons, one light antilepton, plus missing
energy, i.e., $\mu\mu\bar \ell + \E_slash$ ($\ell=e,\mu$) with the
two $\tau$ leptons decaying into the like-sign $\mu$ leptons. The
background is negligible though the signal is hurt by a factor about
$1/36$, the product of the leptonic decay branching ratios of the
$\tau$ lepton.

 From Fig.2, we can see the optimum value of the cross section
 can reach several tens fb, so there could be hundreds of events
 after the signal depressed with the designated integrated luminosity
 above, i.e, $L=500 fb^{-1}$, which may be detected in the future
 ILC experiments.

To see the effect of flavor violating $K_{\tau l}$ on the $\sigma$,
we plot the sigma varying as $K_{\tau l}$ for three values of the
$M_Z'$. We can see from Fig.3 that the cross section $\sigma$ is
larger than $0.1fb$ for $K_{\tau l} \geq 0.4$. Increasing  $K_{\tau
l}$, the maximum value can reach several fb. In this case, there are
about several hundred like-sign $\tau\tau$ production events to be
generated in the future ILC experiments. Considering the rare clear
background of the leptons production, we can still obtain several
events even with small sample of the leptonic decay of the like-sign
$\tau$ leptons.

\begin{figure}[tbh]
\begin{center}
\epsfig{file=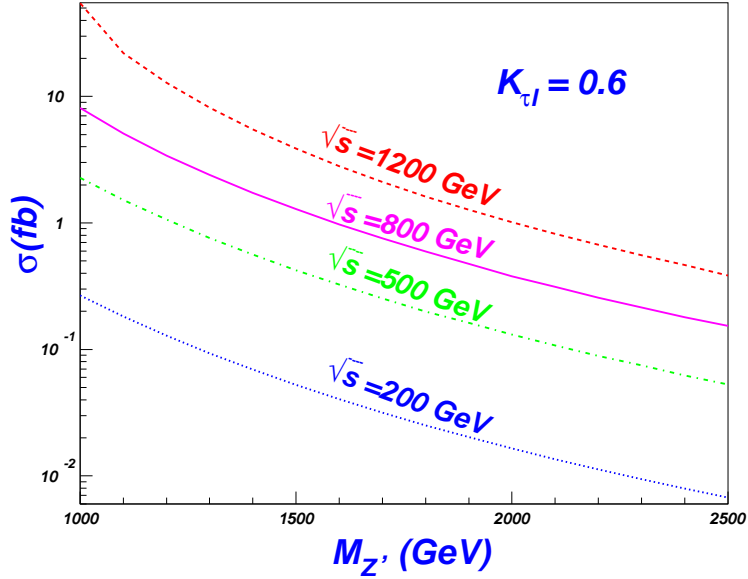,width=10cm} \caption{ The contribution from
top-pion scalars
 $\pi_t^0$ the process $e\gamma \to e^+(\mu^+)\tau\tau$ in TC2
models.} \label{fig2}
\end{center}
\end{figure}

\def\figsubcap#1{\par\noindent\centering\footnotesize(#1)}
\begin{figure}[bht]%
\label{modes}
\begin{center}
\hspace{-0.25cm}
 \parbox{6.05cm}{\epsfig{figure=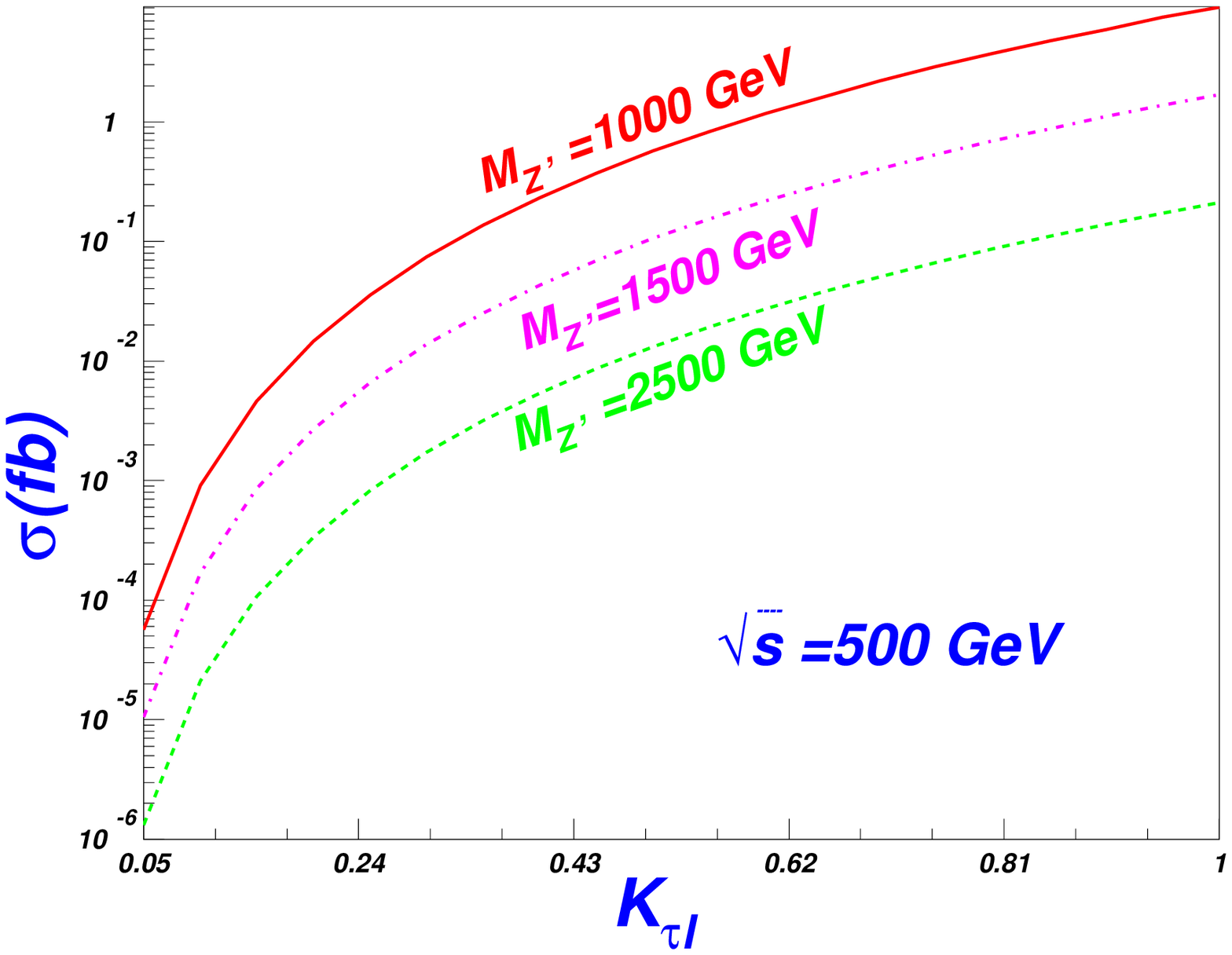,width=6.25cm}
 \figsubcap{a}}
 \hspace*{0.2cm}
 \parbox{6.05cm}{\epsfig{figure=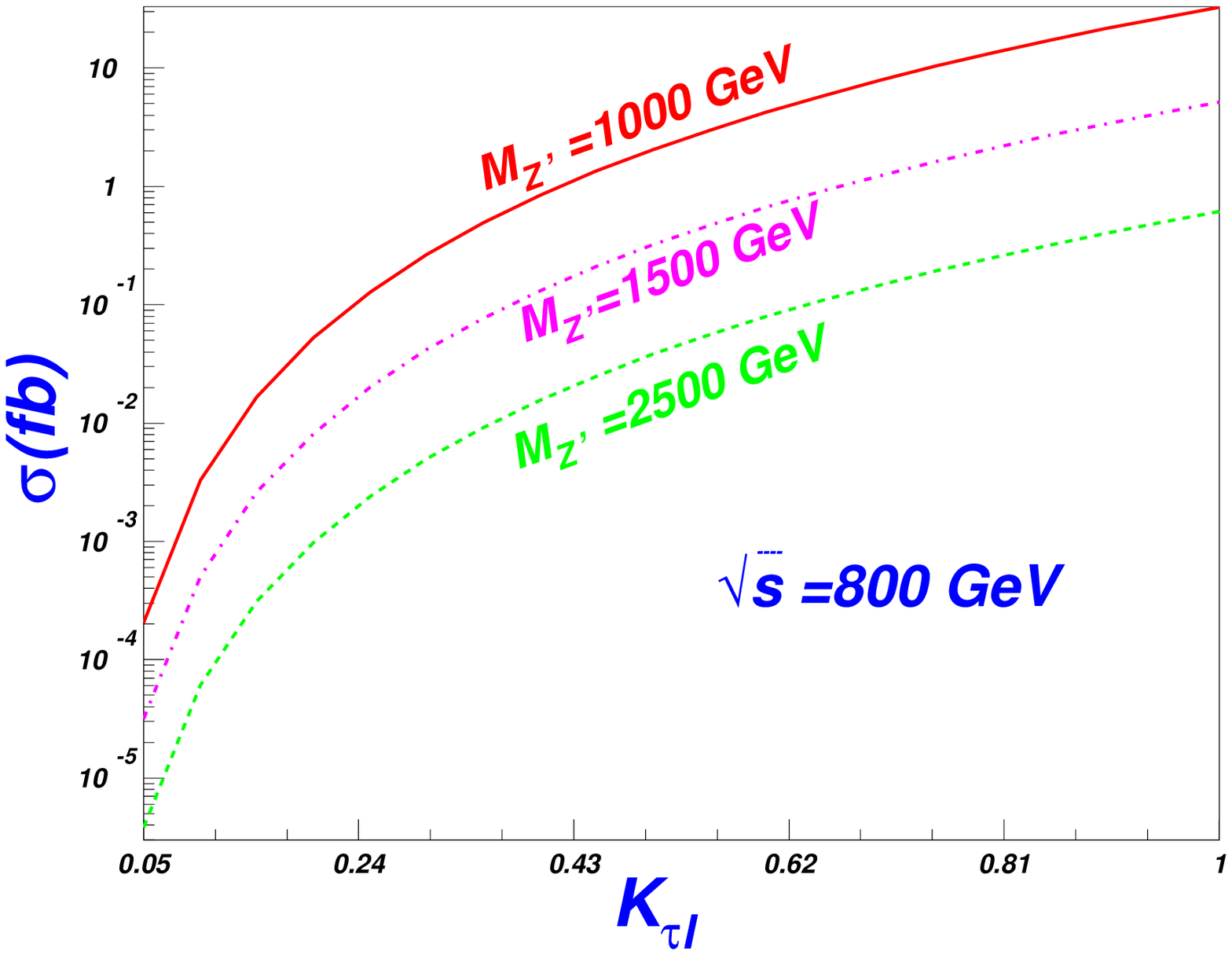,width=6.25cm}
 \figsubcap{b}}
 \caption{The dependence of the cross section $\sigma$ of the LFV process
$e\gamma \to\tau\tau\bar e(\bar\mu)$ on the mixing parameter
$K_{\tau l}$ for  $M_{Z'}=1$, $1.5$, and $2.5$ TeV with (a)
$\sqrt{s}=500$ GeV, (b) $\sqrt{s}=800$ GeV.  \label{fig3} }
\end{center}
\end{figure}

The TC2 models also predict the existence of the neutral state,
top-pion boson $\pi_t^0$, which can also induce the LFV processes
with the couplings:
\begin{eqnarray}
\frac{m_{\tau}}{\nu}
     K_{\tau i}\bar{\tau}\gamma^{5}l_{i}\pi_{t}^{0},
\label{lfv-pi}
\end{eqnarray}
Where $\nu=\nu_{W}/\sqrt{2}\approx174GeV$, $l=\tau,\mu $ or $e$,
$l_{i}$(i=1,2) is the first(second)generation lepton $e$($\mu$), and
$k_{\tau i}$ is the flavor mixing factor between the third-and the
first-or second- generation leptons. There certainly is also the
$FC$ scalar coupling $\pi_{t}^{0}\mu \bar e$. However, Similarly,
the topcolor interactions only contact with the third-generation
fermions, and thus, the flavor mixing between the first- and
second-generation fermions is very small, which can be safely
ignored.
\begin{figure}[tbh]
\begin{center}
\epsfig{file=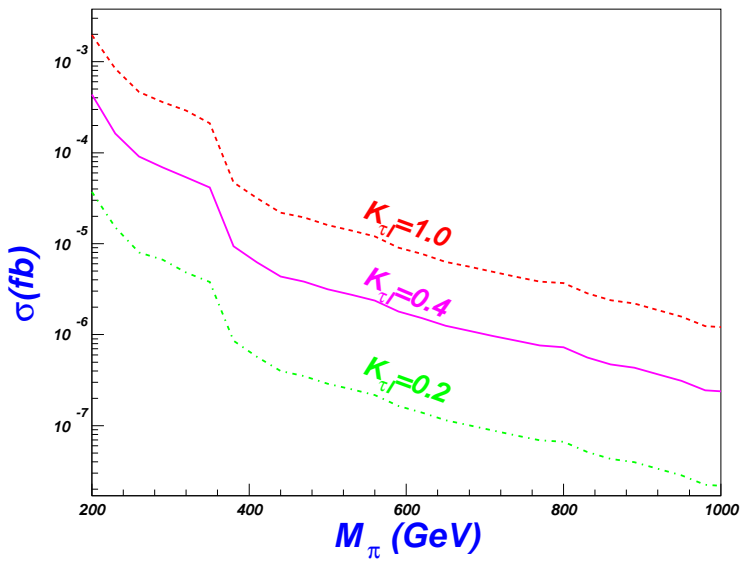,width=10cm} \caption{ The contribution from top-pion scalars
 $\pi_t^0$ the process $e\gamma \to e^+(\mu^+)\tau\tau$ in TC2
models.} \label{fig4}
\end{center}
\end{figure}
We can see from Fig.\ref{fig4} that the cross section $\sigma$ is
smaller than $4\times10^{-3}fb$ for $\sqrt{s}\geq 500$ GeV. The
contribution of $\pi_t^0$ is negligible, which is understandable
from the LFV couplings in eqn.\ref{lfv-pi} since the strengths are
depressed by factor $\frac{m_{\tau}}{\nu}$.

Before ending the discussion, we want to point out that the
like-sign $\tau$ pair productions may be quite unique in probing the
TC2 model at the ILC. To enhance the like-sign $\tau$ pair
production rate to the accessible level at the ILC, the LFV $\tau$
lepton couplings $\tau \bar e Z'$ cannot be too small. The TC2 model
predict sizable tree-level $\tau \bar e Z'$ coupling and thus may
enhance the like-sign $\tau$ pair production rate to the accessible
level at the ILC. In many other popular extensions of the SM, there
are no tree-level $\tau$ lepton LFV couplings and the couplings
$\tau \bar e \phi$ ( $\phi$ is any scalar field) or $\tau \bar e V$
($V=\gamma,Z,g$ or any new gauge boson)  are induced at loop-level,
which are usually too small to make the like-sign $\tau$ pair
productions observable at the ILC. For example, the $\tau$ lepton
LFV couplings are induced at loop-level in the $R$-parity violating
MSSM \cite{rrmutau-susy}. Although they can be much larger than in
the SM, we found that their contribution to the cross sections of $
e \gamma \to \bar e(\bar\mu) \tau\tau $ at the ILC is smaller than
$10^{-5}$ fb.

The search for LFV processes is one of the most interesting
possibilities to test the SM, with the potential for either
 discovering or putting stringent bounds on new physics. In the SM,
 there are no FC coupling at tree-level and at one-loop level they are GIM
suppressed. In models beyond SM, however, new particles may appear
and have significant contributions to the LFV processes. Therefore,
the processes can give an ideal place to search the signals of the
new particles. In this paper, we calculated the contributions of the
gauge boson $Z'$ to the LFV process $e^-\gamma\to
e^+(\mu^+)\tau\tau$ in the framework of TC2 models and discussed the
possiblity of detecting this new particle in the future ILC
experiments. Our numerical results show that the cross section
$\sigma$ induced  by the extra gauge boson $Z'$ is in the range of
the $10^{-1}-1$ fb. In quite
 a large space of the parameters, the cross section $\sigma$ can reach
 servals fb.  So it is possible to detect the signals of the extra gauge boson $Z'$
 via the process $e^-\gamma\to e^+(\mu^+)\tau\tau$ at the
 $e\gamma$ colliders based on the ILC experiments.

\vspace{2.5mm}
 {\bf \large Acknowledgement}
\vspace{2.5mm}

  We would like to thank J. J. Cao, J. M. Yang
and C. P. Yuan for helpful discussion.

\end{document}